\documentclass[aps,pra,superscriptaddress,showpacs,twocolumn,nofootinbib,10pt]{revtex4-1}

\usepackage{amsfonts,amssymb,amsmath}
\usepackage{graphics,graphicx,epsfig}
\usepackage{amsthm}
\usepackage{epstopdf}

\def\identity{\leavevmode\hbox{\small1\kern-3.8pt\normalsize1}}

\newcommand{\ket}[1]{\left | #1 \right\rangle}
\newcommand{\bra}[1]{\left \langle #1 \right |}

\newcommand{\ketbra}[2]{| #1 \rangle \langle #2|}
\newcommand{\proj}[1]{\ket{#1}\bra{#1}}
\renewcommand{\epsilon}{\varepsilon}

\begin{document}

\title{Optimized state independent entanglement detection based on geometrical threshold criterion}

\author{Wies\l aw~Laskowski}

\affiliation{Institute of Theoretical Physics and Astrophysics, University of Gda\'nsk, PL-80-952 Gda\'nsk, Poland}

\author{Christian~Schwemmer}
\author{Daniel~Richart}

\affiliation{Max-Planck-Institut f\"ur Quantenoptik, Hans-Kopfermann-Strasse 1, D-85748 Garching, Germany}
\affiliation{Department f\"ur Physik, Ludwig-Maximilians-Universit\"at, D-80797 M\"unchen, Germany}

\author{Lukas~Knips}

\affiliation{Max-Planck-Institut f\"ur Quantenoptik, Hans-Kopfermann-Strasse 1, D-85748 Garching, Germany}
\affiliation{Department f\"ur Physik, Ludwig-Maximilians-Universit\"at, D-80797 M\"unchen, Germany}

\author{Tomasz~Paterek}

\affiliation{School of Physical and Mathematical Sciences, Nanyang Technological University, Singapore}
\affiliation{Centre for Quantum Technologies, National University of Singapore, Singapore}

\author{Harald~Weinfurter}

\affiliation{Max-Planck-Institut f\"ur Quantenoptik, Hans-Kopfermann-Strasse 1, D-85748 Garching, Germany}
\affiliation{Department f\"ur Physik, Ludwig-Maximilians-Universit\"at, D-80797 M\"unchen, Germany}

\begin{abstract}
Experimental procedures are presented for the rapid detection of entanglement of unknown arbitrary quantum states.
The methods are based on the entanglement criterion using accessible correlations and the principle of correlation complementarity.
Our first scheme essentially establishes the Schmidt decomposition for pure states, with few measurements only and without the need for shared reference frames.
The second scheme employs a decision tree to speed up entanglement detection.
We analyze the performance of the methods using numerical simulations and verify them experimentally for various states of two, three and four qubits.
\end{abstract}

\pacs{03.67.Mn}

\maketitle

\section{Introduction}

Entanglement is one of the most fundamental features of quantum physics and is considered as the key resource for quantum information processing \cite{HORODECCY, PAN, NIELSEN}. In order to detect entanglement highly efficient witness operators are widely used nowadays \cite{WITNESS,BOURENNANE_ETAL_2004,TERHAL2000,LEWENSTEIN_ETAT_2000,BRUSS_ETAL_2002,GH2003,GT2009}. However, these operators give conclusive answers only for states close to the target state. To detect entanglement of arbitrary states, positive, but not completely positive maps \cite{WITNESS,HORODECCY}, are the most universal entanglement identifiers. However, they are laborious to use as they require full state tomography. Therefore, more efficient schemes to detect entanglement are most wanted.

It has been recently shown that the presence of entanglement in a quantum state is fully characterized by suitable combinations of experimentally accessible correlations and expectation values of local measurements \cite{FRIENDLY}. This enables a simple and practical method to reveal entanglement of all pure states and some mixed states by measuring only few correlations \cite{LASKOWSKI}. Since the method is adaptive it does not require a priori knowledge of the state nor a shared reference frame between the possibly remote observers and thus greatly simplifies the practical application.

Here we extend these results and analyze in detail the possible performance of two schemes for entanglement detection. The first one can be seen as an experimental implementation of Schmidt decomposition, which identifies the maximal correlations through local measurements only. The second scheme shows how to deduce a strategy (decision tree) to find the maximal correlations of an unknown state and obtain a rapid violation of the threshold identifying entanglement even for an arbitrary number of qubits. 
The physical principle behind both of our schemes is correlation complementarity~\cite{CORRELATION_COMPLEMENTARITY}.
It makes use of trade-offs between correlations present in quantum states. Once a measured correlation is big other related correlations have to be small and it is advantageous to move to measurements of the remaining correlations. This simplifies the entanglement detection scheme as a lower number of correlation measurements is required.

\section{Entanglement criterion}
\label{SEC_CRITERION}

A quantum state is entangled if the sum of squared measured correlations exceeds a certain bound \cite{FRIENDLY}.
This identifier thus neither requires the measurement of all correlations in a quantum state nor the reconstruction of the density matrix.
Rather, it is now the goal to find strategies that minimize the number of correlation measurements. 
We show how this can be done in different ways described in the subsequent sections.
The first method is to identify a Schmidt decomposition from local results and filtering when necessary,
the second is a particularly designed decision tree based on correlation complementarity.

Any $N$ qubit density matrix can be expressed as:
\begin{equation}
\rho = \frac{1}{2^N} \sum_{\mu_1,...,\mu_N=0}^3 T_{\mu_1...\mu_N}
\sigma_{\mu_1} \otimes ... \otimes \sigma_{\mu_N},
\label{STATE}
\end{equation}
where $\sigma_{\mu_n} \in \{\sigma_0,\sigma_x,\sigma_y,\sigma_z\}$ is the
respective local Pauli operator of the $n$th party ($\sigma_0$ being the identity matrix) and the real
coefficients $T_{\mu_1...\mu_N} \in [-1,1]$ are the
components of the correlation tensor $\hat T$. They
are given by the expectation values of the products of local Pauli
observables, $T_{\mu_1...\mu_N} = \mbox{Tr}[\rho (\sigma_{\mu_1} \otimes ... \otimes \sigma_{\mu_N})]$, and can be determined by local measurements performed on each qubit. 

For $N$ qubit states, pure or mixed, the following sufficient condition for entanglement holds~\cite{FRIENDLY}:
\begin{equation}
\sum_{i_1, \dots, i_N = 1}^3 T_{i_1\dots i_N}^2 > 1 \quad \Rightarrow \quad \rho \textrm{ is entangled}.
\label{SIMPLE_CRIT}
\end{equation}
Note, to prove that a state is entangled, it is sufficient to break the threshold, i.e., in general it is not necessary to measure all correlations. 
Using fundamental properties of the correlation tensor, we design schemes to minimize the number of required correlation measurements.

\section{Schmidt decomposition}
\label{SEC_SCHMIDT}

Any pure state of two qubits admits a Schmidt decomposition \cite{SCHMIDT, PERES}
\begin{equation}
\ket{\psi_S} = \cos \theta \ket{a} \ket{b} + \sin \theta \ket{a_\perp} \ket{b_\perp}, \quad \theta \in [0,\tfrac{\pi}{4}].
\label{SCHMIDT}
\end{equation}
where the local bases $\{\ket{a},\ket{a_\perp}\}$ and $\{\ket{b}, \ket{b_\perp}\}$ are called the Schmidt bases of Alice and Bob.

This is an elementary description of bipartite pure quantum states, where the existence of a second term in the decomposition directly indicates entanglement. 
In addition, in the Schmidt bases, the correlation tensor of a two-qubit state takes a particularly simple form and shows maximal correlations in the state.
Therefore, finding the Schmidt bases can be regarded as a redefinition of the measuring operators relative to the state and thus leads to rapid entanglement detection, in at most three subsequent measurements of correlations.

Once the bases are known, Alice constructs her local measurements $\sigma_{z'} = \proj{a} - \proj{a_\perp}$ and $\sigma_{y'} = i \ket{a_\perp} \bra{a} - i \ket{a} \bra{a_\perp}$, and so does Bob.
They can now detect entanglement by using the simple criterion (\ref{SIMPLE_CRIT}) with only two correlation measurements because $T_{z'z'}^2 + T_{y'y'}^2 = 1 + \sin^2 2 \theta > 1$ for all pure entangled states.
Note that since the bases of Alice and Bob are determined on the fly the laboratories are not required to share a common reference frame.

In the next sections we present how to find the Schmidt bases (up to a global phase) from the experimental results gathered on individual qubits.
We split the discussion into two cases, of non-vanishing and vanishing Bloch vectors, i.e. local averages $(T_{x0},T_{y0},T_{z0})$, describing the states of the individual qubits.

\begin{figure}
\includegraphics[width=0.44\textwidth]{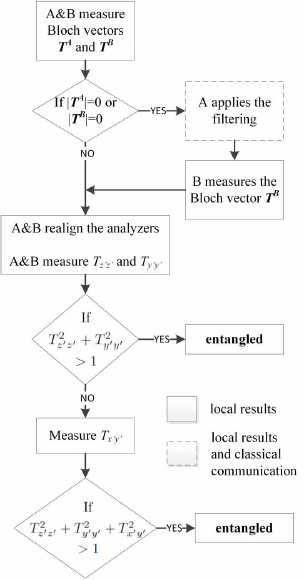}
\caption{\label{FIG_ALGO} The systematic way to experimentally verify entanglement of arbitrary pure two-qubit state without any a priori knowledge and in the absence of common reference frame. The steps of this diagram are described in detail in the corresponding sections of the main text.}
\end{figure}

This systematic procedure to verify entanglement in a pure two-qubit state is represented in Fig.~\ref{FIG_ALGO}, with the sections describing the particular steps.

\subsection{From non-vanishing Bloch vectors \\ to Schmidt bases}

\label{localresults}

Consider first the case of non-zero Bloch vectors. The Schmidt bases of Alice and Bob are related to the standard bases as follows:
\begin{eqnarray}
\ket{a} & = & \cos \xi_{A} \ket{0} + e^{i \varphi_{A}} \sin \xi_{A} \ket{1}, \label{schmidtAlice} \nonumber \\
\ket{a_\perp} & = & \sin \xi_{A} \ket{0} - e^{i \varphi_{A}} \cos \xi_{A} \ket{1}, \label{schmidtAliceP} \nonumber\\
\ket{b} & = & \cos \xi_{B} \ket{0} + e^{i \varphi_{B}} \sin \xi_{B} \ket{1}, \label{schmidtBob} \nonumber \\
\ket{b_\perp} & = & e^{i \delta}(\sin \xi_{B} \ket{0} - e^{i \varphi_{B}} \cos \xi_{B} \ket{1}). \label{schmidtBobP}
\end{eqnarray}
The global phase of $\ket{b_\perp}$ is relevant and required for the characterization of an arbitrary pure state, as can be seen from parameter counting. 
An arbitrary pure two-qubit state is parametrized by six real numbers (four complex amplitudes minus normalization condition and an irrelevant global phase).
Plugging Eqs.~(\ref{schmidtBobP}) into the Schmidt decomposition (\ref{SCHMIDT}) we indeed find the relevant six real parameters.

Any two-qubit state written in the standard bases of Alice and Bob can be brought into the Schmidt basis of Alice by the transformation
\begin{eqnarray}
U(\xi_A,\varphi_A)
& = & \ketbra{0}{a} + \ketbra{1}{a_\perp} \nonumber \\
& = & \cos \xi_A \ketbra{0}{0}+ e^{- i \varphi_A} \sin \xi_A \ketbra{0}{1} \nonumber \\
&+& \sin \xi_A \ketbra{1}{0}  - e^{-i \varphi_A} \cos \xi_A \ketbra{1}{1}.
\end{eqnarray}
The coefficients $\xi_A$ and $\varphi_A$ of this transformation can be read from a non-vanishing normalized Bloch vector:
\begin{equation}
\vec \alpha \equiv \frac{\vec T^{A}}{|\vec T^{A}|} = (\sin 2 \xi_A \cos \varphi_A, \sin 2 \xi_A \sin \varphi_A, \cos 2 \xi_A).
\end{equation}
Finally, the coefficients of the Schmidt basis in the standard basis
are functions of components of vector $\vec \alpha = (T_{x0},T_{y0},T_{z0}) / \sqrt{T_{x0}^2+T_{y0}^2+T_{z0}^2}$
built out of experimentally accessible, local expectation values of Pauli measurements:
\begin{eqnarray}
\cos \xi_A = \sqrt{\frac{1+{\alpha_{z}}}{2}}, & \quad & \sin \xi_A = \sqrt{\frac{1- {\alpha_{z}}}{2}}, \nonumber \\
\cos \varphi_A = \frac{{\alpha_{x}}}{\sqrt{1 - \alpha_{z}^2}}, & \quad & \sin \varphi_A = \frac{{\alpha_{y}}}{\sqrt{1 - \alpha_{z}^2}}.
\end{eqnarray}
If ${T_{z0}} = \pm 1$, the standard basis is the Schmidt basis. 
Note that instead of transforming the state we can as well transform the measurement operators $\sigma_{n'}  = U^{\dagger} \sigma_n U$. 
The new operators are given by:
\begin{eqnarray}
\sigma_{x'}&=& \frac{-\alpha_x \alpha_z \sigma_x - \alpha_y \alpha_z \sigma_y + (1-\alpha_z^2) \sigma_z}{\sqrt{1-\alpha_z^2}}, \nonumber \\
\sigma_{y'} &=& \frac{{\alpha_{y} \sigma_x + \alpha_{x} \sigma_y}}{\sqrt{1 - \alpha_{z}^2}}, \nonumber \\
\sigma_{z'} &=& \alpha_{x} \sigma_x + \alpha_{y} \sigma_y + \alpha_{z} \sigma_z,
\label{new_directions}
\end{eqnarray}
and the Schmidt basis is the $z'$ basis, i.e. $\sigma_{z'} = \proj{a} - \proj{a_\perp}$.

The equivalent analysis has to be done for the Schmidt basis of Bob.
In summary, the Schmidt bases are defined by the direction of the Bloch vectors of reduced states, up to a global phase.
The global phase of $\ket{b_\perp}$ shows up as a relative phase in the Schmidt decomposition.
As it is not obtainable by local measurements it influences entanglement detection using operators (\ref{new_directions}).

\subsection{Entanglement detection}
\label{SEC_ENT_DET}

Let us denote the basis established by local measurements of Bob by $\{|\tilde b \rangle, |\tilde b_\perp \rangle \}$, i.e. $|b \rangle= |\tilde b \rangle   $ and $|b_\perp \rangle=e^{i \delta} |\tilde b_\perp \rangle$.
Using the locally determined bases the Schmidt decomposition takes the form
\begin{equation}
\ket{\psi_S} = \cos \theta \ket{a} | \tilde b \rangle + e^{i \delta} \sin \theta \ket{a_\perp} |\tilde b_\perp \rangle.
\end{equation}
The correlations that Alice and Bob observe in the measurements related to locally determined bases are
$T_{z'z'} =1$ and $T_{x'x'} = \sin 2\theta \cos\delta$, $T_{y'y'} = -\sin 2\theta \cos\delta$, $T_{x'y'} = \sin 2\theta \sin\delta$, $T_{y'x'} = \sin 2\theta \sin\delta$.
Note that the correlation $T_{y'y'}$ would vanish for $\cos\delta=0$ and the two measurements $T_{z'z'}$ and $T_{y'y'}$ are not sufficient any more (they were sufficient if the full knowledge about the Schmidt bases had been available).
In such a case, however, the other two correlations, $T_{x'y'}$ and $T_{y'x'}$, are non-zero, and can be used to reveal entanglement. 
If the first two measurements are not sufficient to overcome the entanglement  threshold of~(\ref{SIMPLE_CRIT}),
the third measurement of $T_{x'y'}$ correlations will definitely allow exceeding the threshold for every pure entangled state.

\subsection{Vanishing Bloch vectors. Filtering}
\label{VanishingBlochvectors}

If the two-qubit state is maximally entangled, i.e. in the Schmidt decomposition $\ket{\kappa} = \frac{1}{\sqrt{2}}( \ket{a} \ket{b} + \ket{a_\perp} \ket{b_\perp})$, 
the Bloch vector is of zero length, $|\vec T^A| = 0$, and the whole system admits infinitely many Schmidt decompositions.
For every unitary operation of Alice, $U$, there exists an operation of Bob, $U'$, such that the state is unchanged:
\begin{equation}
U \otimes U' \ket{\kappa} = \ket{\kappa}.
\end{equation}
Therefore, any basis of, say,  Alice can serve as the Schmidt basis as soon as we accordingly update the basis of Bob.
Our strategy to reveal the corresponding basis of Bob is to filter in the chosen Schmidt basis of Alice.
It is best to explain it on an example.
Assume Alice chooses the standard basis as her Schmidt basis.
Due to the mentioned invariance of the maximally entangled state
there exists a Schmidt basis of Bob such that
\begin{equation}
\ket{\kappa} = \frac{1}{\sqrt{2}}(\ket{0} \ket{b'} + \ket{1} \ket{b_\perp'}).
\end{equation}
The basis of Bob can be found by filtering of Alice $F = \epsilon \ket{0} \bra{0} + \ket{1} \bra{1}$ with $\epsilon\in [0,1)$.
We implemented this operation experimentally and provide the details later.
In short, we use devices which are transparent to the $\ket{1}$ state,
but probabilistically ``reflect'' the $\ket{0}$ state.
If we imagine a perfect detector is observing a port of this device where the reflected particle travels,
and we see no detection, the filter operation is performed on the initial state.
If Alice applies the filtering on her qubit and informs Bob that the filtering was successful 
the initial state becomes
\begin{equation}
(F \otimes \openone) \ket{\kappa} \to \frac{1}{\sqrt{1 + \epsilon^2}} (\epsilon \ket{0} \ket{b'} + \ket{1} \ket{b_\perp'}).
\end{equation}
The result of filtering is that for Bob a Bloch vector emerges and we can again use the method described above to find his Schmidt basis.

\subsection{Performance}

Summing up all required steps we see that to experimentally verify entanglement of any pure two-qubit state without any further a priori knowledge requires at least $2 \times 3$ local measurements to determine the Schmidt bases and sometimes filtering requiring 3 local measurements more. 
Finally two more (or three if $\delta = \pi/2$) correlation measurements allow to verify the entanglement criterion.

\begin{figure}
\includegraphics[width=0.3\textwidth]{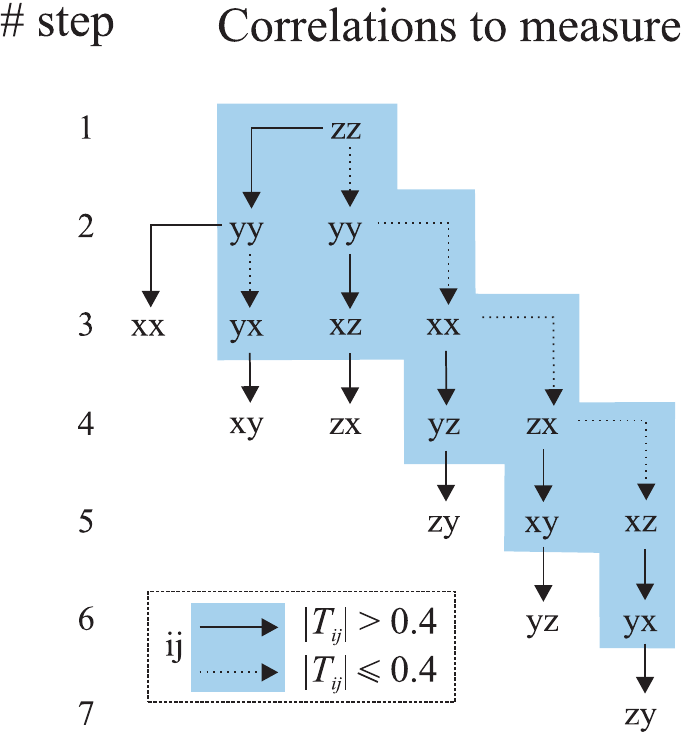}
\caption{Decision strategies to detect entanglement. Start with a measurement of the correlation $T_{zz}$ and proceed with the correlation along the solid (dotted) arrow if the measured correlation is higher (lower) than the chosen threshold value; here of $t=0.4$. Due to correlation complementarity there is a good chance of detecting entanglement in a small number of steps.
The measurements in the blue shaded area suffice to detect all maximally entangled pure states with Schmidt-basis vectors $x,y$ or $z$.}
\label{FIG_DT}
\end{figure}

\section{Decision Tree}
\label{section_DT}

Our second algorithm for entanglement detection does not even require any initial measurements and directly applies also to mixed states.
We will split the presentation into bipartite and multipartite cases. The decision tree provides an adaptive method to infer the next measurement setting from previous results. 

\subsection{Two qubits}
\label{SEC_2Q}

Alice and Bob choose three orthogonal local directions $x,y$ and $z$ independently from each other and agree to only measure correlations along these directions.
In Fig. \ref{FIG_DT} we show exemplarily which correlations should be measured in order to detect entanglement in a small number of steps.
Starting with a measurement of $T_{zz}$, one continues along the solid (dotted) arrow, if the correlation is higher (lower) than some threshold value $t$.
We performed detailed numerical analysis on how the efficiency of entanglement detection depends on the threshold value.
The efficiency is quantified by the percentage of entangled states detected at various steps of the decision tree.
It turns out that the efficiency does not depend much on the threshold value and the best results are obtained for $t = 0.4$.
We therefore set this threshold value in all our simulations.

The construction of the tree is based on the principle of correlation complementarity \cite{CORRELATION_COMPLEMENTARITY,TG2005, WW_CC_2008, WW_CC_2010}:
in quantum mechanics there exist trade-offs for the knowledge of dichotomic observables with corresponding anti-commuting operators.
For this reason, if the correlation $|T_{zz}|$ is big, correlations $|T_{zx}|,|T_{zy}|,|T_{xz}|$ and $|T_{yz}|$ have to be small because their corresponding operators anti-commute with the operator $\sigma_{z} \otimes \sigma_{z}$.
Therefore, the next significant correlations have to lie in the $xy$ plane of the correlation tensor and the next step in the tree is to measure the $T_{yy}$ correlation.

In cases in which going through the whole tree did not reveal entanglement 
we augmented it with additional measurements of correlations that were not established until that moment. The order of the additional measurements also results from the correlation complementarity.
With every remaining measurement we associate the ``priority'' parameter
\begin{equation}
P_{ij} = \sum_{k \ne i} P_{ij}(T_{kj}) + \sum_{l \ne j} P_{ij}(T_{il}),
\end{equation}
that depends on the measurements of the decision tree in the following way
\begin{equation}
P_{ij}(T_{mn}) = \Big\{
\begin{array}{rl}
T_{mn}^2 & \textrm{  if } T_{mn} \textrm{ was performed before,} \\
0 & \textrm{  else.}
\end{array}
\end{equation}
According to the correlation complementarity if the value of the corresponding parameter is small there is a bigger chance that this correlation is significant.
Therefore, the correlations $T_{ij}$ with lower values of $P_{ij}$ are measured first. 

Let us illustrate this on the following example. 
The measured correlations of the decision tree are as follows: $T_{zz}=0.7$, $T_{yy}=0.6$, and $T_{xx}=0.1$. 
Therefore, $P_{xy} = P_{yx} = T_{xx}^2 + T_{yy}^2 = 0.37$ , $P_{xz} = P_{zx} = T_{xx}^2 + T_{zz}^2 = 0.5$, and $P_{yz} = P_{zy} = T_{zz}^2 + T_{yy}^2 = 0.85$.
Accordingly, the order of the remaining measurements is as follows: 
first measure $xy$, then $yx$, next $xz,zx,yz$ and $zy$.

\subsection{Many qubits}
\label{SEC_MANY_QUBITS}

Correlation complementarity, which holds also in the multipartite case, states that for a set $\{\alpha_1,\dots, \alpha_k \}$ of dichotomic mutually anti-commuting multiparty operators 
the following trade-off relation is satisfied by all physical states:
\begin{equation}
T_{\alpha_1}^2 + \dots + T_{\alpha_k}^2 \le 1,
\end{equation}
where $T_{\alpha_1}$ is the expectation value of observable $\alpha_1$ and so on.
Therefore, if one of the expectation values is maximal, say $T_{\alpha_1} = \pm 1$, the other anti-commuting observables have vanishing expectation values and do not have to be measured. 
In this way we exclude exponentially many, in the number of qubits, potential measurements because that many operators anti-commute with $\alpha_1$,
and we apply correlation complementarity pairwise to $\alpha_1$ and one of the anti-commuting operators.
This motivates taking only sets of commuting operators along the branches of the decision tree that should be followed if the measured correlations are big.

We are thus led to propose the following algorithm generating one branch of the decision tree 
in which the first measurement, called $X \otimes X \otimes \dots \otimes X$, is assumed to have a big expectation value.

\begin{itemize}

\item[(i)] Generate all $N$-partite Pauli operators that commute with $X \otimes X \otimes \dots \otimes X$.

Such operators have an even number of local Pauli operators different than $X$.
Accordingly, their number is given by: $\sum_{j=1}^{\lfloor \frac{N}{2} \rfloor} 2^{2j} {N \choose 2j} = \frac{1}{2}(3^N-1) - \mathrm{Odd}(N)$,
where $\mathrm{Odd}(N) = 1$ if $N$ is odd, and $0$ otherwise. For example, in the three qubit case the set of operators commuting with $XXX$ consists of 12 operators: $XZZ$, $ZZX$, $ZXZ$, $XYY$, $YYX$, $YXY$, $XYZ$, $XZY$, $YXZ$, $YZX$, $ZXY$, $ZYX$.  

\item[(ii)] Group them in strings of mutually commuting operators that contain as many elements as possible.

We verified for $N$ up to eight qubits (and conjecture in general)
that the length of the string of mutually commuting operators is $L = 2^{N-1} + \mathrm{Even}(N)$,
where $\mathrm{Even}(N) = 1$ if $N$ is even, and $0$ otherwise. In our three qubit example, we have the following strings: 
$\{XXX$, $YXZ$, $XZZ$, $YZX\}$, 
$\{XXX$, $YYX$, $XYZ$, $YXZ\}$.
$\{XXX$, $XZY$, $YZX$, $YXY\}$, 
$\{XXX$, $XZZ$, $ZXZ$, $ZZX\}$, 
$\{XXX$, $XYZ$, $ZXZ$, $ZYX\}$, 
$\{XXX$, $YXY$, $YYX$, $XYY\}$, 
$\{XXX$, $XYY$, $ZXY$, $ZYX\}$, 
$\{XXX$, $ZZX$, $XZY$, $ZXY\}$,

We denote the number of such strings by $M$.

\item[(iii)] Arrange the operators within the strings and sort the strings such that they are ordered with the same operator in the first position, then, if possible, second, third, etc.

In this way we produce a set of strings $\{S_1,S_2, \dots, S_M\}$ such that in the first position of every string we have $S_{j,1} = X \otimes X \otimes \dots \otimes X$,
in the second position the number of different operators is smaller or equal to the number of different operators in the third position etc.  After applying that operation in the three qubit example, we obtain:  
$\{XXX$, $XZZ$, $ZXZ$, $ZZX\}$, 
$\{XXX$, $XZZ$, $YXZ$, $YZX\}$, 
$\{XXX$, $XZY$, $YZX$, $YXY\}$, 
$\{XXX$, $XZY$, $ZZX$, $ZXY\}$, 
$\{XXX$, $XYZ$, $YXZ$, $YYX\}$, 
$\{XXX$, $XYZ$, $ZXZ$, $ZYX\}$, 
$\{XXX$, $XYY$, $YXY$, $YYX\}$, 
$\{XXX$, $XYY$, $ZXY$, $ZYX\}$.

\item[(iv)] Connect the operators of the string $S_1$ with continuous arrows:
\begin{equation}
S_{1,1} \longrightarrow S_{1,2} \longrightarrow \dots \longrightarrow S_{1,L}.
\end{equation}

\item[(v)] For all other strings $S_j$, with $j=2,\dots,M$, check on which position string $S_j$ differs from $S_{j-1}$. Let us denote this position by $d$. 
At these positions strings can be connected with another type of arrows, yielding the tree as
\begin{equation}
S_{j-1,d}  \dashrightarrow S_{j,d} \longrightarrow S_{j,d+1} \longrightarrow \dots \longrightarrow S_{j,L}.
\end{equation}

\end{itemize}

\begin{figure}
\includegraphics[width=0.46\textwidth]{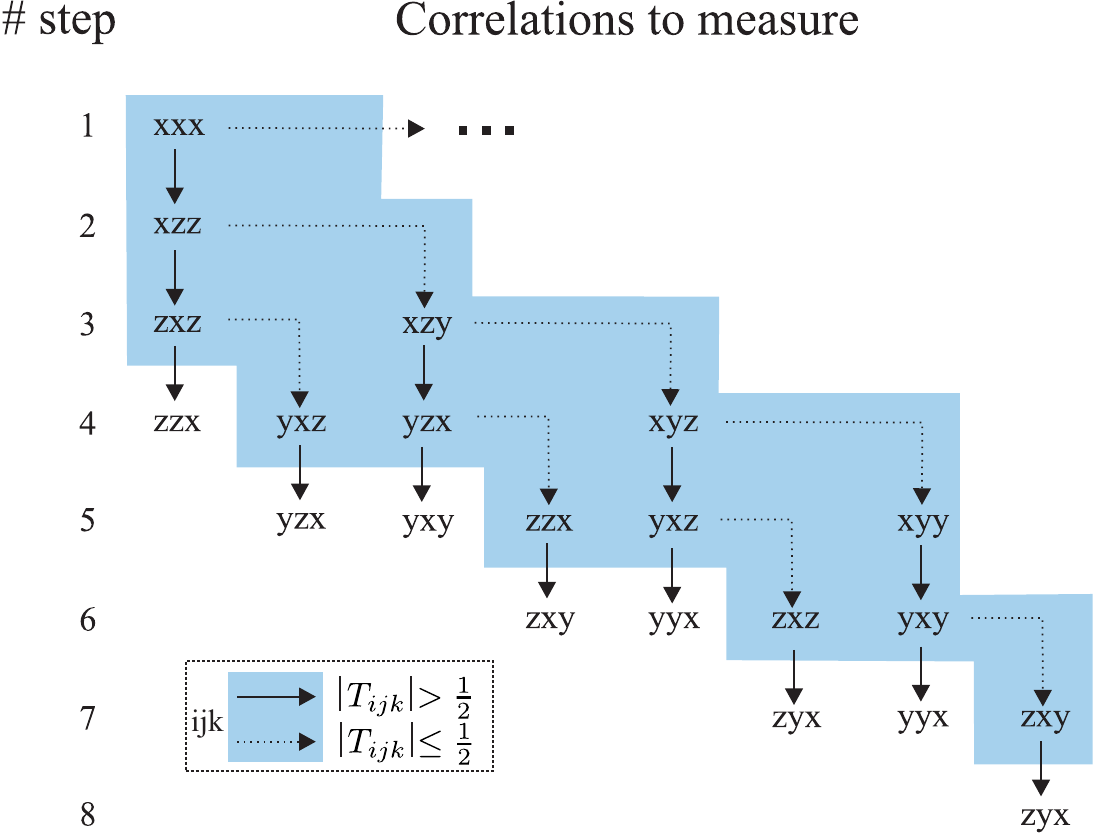}
\caption{One branch of the decision tree for three qubits, starting with a measurement of the correlation $T_{xxx}$ assumed to be big. 
The best efficiency is obtained for the threshold $t = 0.5$.}
\label{DT_3QUBITS}
\end{figure}

With the strings of operators from step (iii) and choosing some threshold whether to follow one or the other string we can now build up a decision tree as shown in Fig. \ref{DT_3QUBITS}. 
Its essential feature is that an operator with big expectation value is followed only by the measurements of commuting operators, irrespectively of their expectation values.

\subsection{Bloch correlations}

Finally, it would be useful to establish a measurement suitable as a starting point of the decision tree, i.e. such that the measured correlations have a good chance of being big.
A natural candidate is to connect both methods discussed here and check whether the correlation measured along the Bloch vectors of every observer (we denote it as Bloch correlations) gives values close to the maximal correlation in a pure state.
We verified this numerically and found that the Bloch correlations are larger than $\frac{3}{4}$ of maximal correlations in a pure state in $100\%$ of two-qubit states, 
$69\%$ of three-qubit states, but only in $27\%$ of four-qubit states and $3\%$ of five-qubit states.
Therefore, the Bloch correlations give a very good starting point of the decision tree only for two and three qubits. 
We leave it as an open question whether a simple and reliable method exists that identifies the maximal correlations of a pure multi-qubit state.

\subsection{Performance}

Let us analyze the results on the entanglement detection efficiency for different classes of two qubit states.
As explained in section \ref{SEC_ENT_DET}, at maximum three correlation measurements are sufficient to detect entanglement once the local Schmidt bases of Alice and Bob are known. 
Here, in contrast, we study how many correlation measurements are needed when the decision tree is applied to an unknown entangled state.  

\begin{figure}
\includegraphics[width=0.48\textwidth]{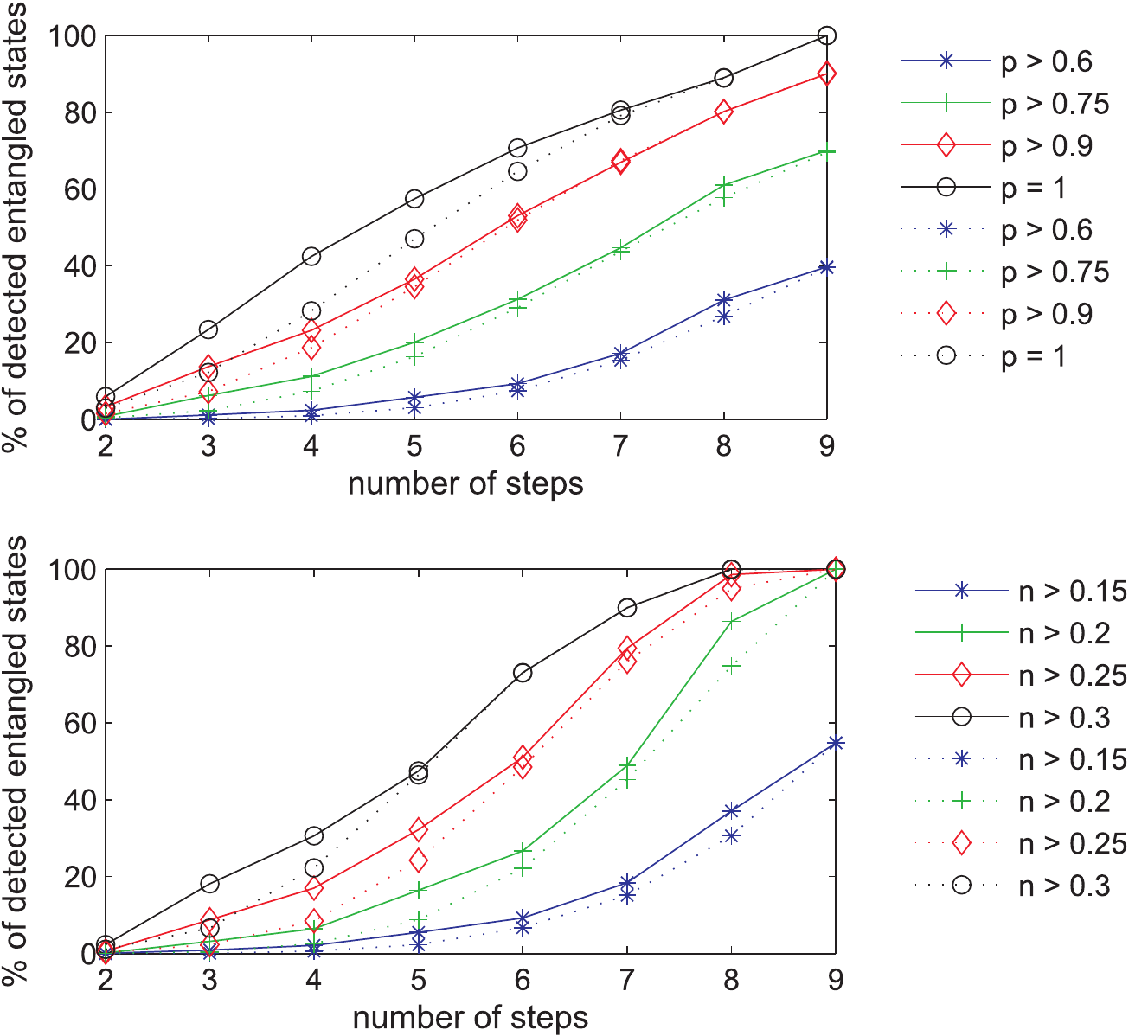}
\caption{Efficiency of the decision tree for two-qubit random mixed states.
The states were uniformly sampled according to the Haar measure.
The efficiency increases with the purity of the state (top panel) as well as with the amount of entanglement in a tested state (bottom panel).
Note that all pure entangled states are detected after nine steps as well as all the states with negativity more than $0.2$ independently of their purity.
Solid lines show the results when using the decision tree (DT), dotted lines when using random choices for the measurements.
}
\label{SI_FIG_RANDOM_MIXED}
\end{figure}

The dependence of the efficiency of the algorithm on  the number of steps involved can be seen in Fig. \ref{SI_FIG_RANDOM_MIXED}. The efficiency is defined by the fraction of detected entangled states with respect to all randomly generated entangled states. For nine steps the algorithm detects all pure entangled states. This is expected because Eq. (\ref{SIMPLE_CRIT}) is a necessary and sufficient condition for entanglement in the case of pure states.
In the case of mixed states, Fig. \ref{SI_FIG_RANDOM_MIXED} shows how the efficiency of the algorithm scales with the purity of the tested state.
Since condition (\ref{SIMPLE_CRIT}) is similar to the purity of a state, obviously, the scheme succeeds the faster the more pure a state is.

Fig. \ref{SI_FIG_RANDOM_MIXED} also shows that the efficiency of the decision tree grows with the amount of entanglement in a state as characterized by the negativity \cite{NEGATIVITY}. It turns out that all the states are detected by the tree that have negativity more than $\frac{1}{5}$.

We also compared the efficiency of the decision tree algorithm to entanglement detection based on a random order of measurements. In the first step of this protocol Alice and Bob randomly choose one of 9 measurements that also enter the decision tree. In the second step they randomly measure one of the 8 remaining measurements and so on. At each step condition (\ref{SIMPLE_CRIT}) is checked for entanglement detection. Of course the two methods converge for higher number of measurements. For small number of measurements the decision tree detects entanglement roughly one step faster than the random measurement method. The advantage of the decision tree with respect to a random choice of the correlations is more pronounced for a higher number of qubits (see section~\ref{SEC_3Q}).

Condition (\ref{SIMPLE_CRIT}) alone, i.e. without considering specific, state dependent metrics (see \cite{FRIENDLY}) cannot detect all mixed entangled states. As an illustration of how the decision tree works for mixed states we first consider Werner states.
It turns out that not all entangled states of this family can be detected whereas the following example shows a family of mixed states for which all the states are detected.

\subsubsection{Werner states}

Consider the family of states
\begin{equation}
\rho = p | \psi^- \rangle \langle \psi^- | + (1-p) \frac{1}{4} \openone,
\label{WERNER_STATE}
\end{equation}
where $\ket{\psi^-} = \frac{1}{\sqrt{2}}(\ket{01} - \ket{10})$ is the Bell singlet state,
$\frac{1}{4} \openone$ describes the completely mixed state (white noise), and $p$ is a probability \cite{WERNER}.
Its correlation tensor, written in the same coordinate system for Alice and Bob, is diagonal with entries $T_{xx} = T_{yy} = T_{zz} = -p$, arising from the contribution of the entangled state.
The states (\ref{WERNER_STATE}) are entangled if and only if $p > \frac{1}{3}$,
whereas the decision tree reveals the entanglement only for $p > \frac{1}{\sqrt{3}} \approx 0.577$.

\subsubsection{Entanglement mixed with colored noise}

An exemplary class of density operators for which the decision tree detects all entangled states is provided by:
\begin{equation}
\gamma = p | \psi^- \rangle \langle \psi^- | + (1-p) \proj{01},
\end{equation}
i.e. the maximally entangled state is mixed with colored noise $\ket{01}$ bringing anti-correlations along the local $z$ axes.
For this case, quite common for type-II parametric down-conversion sources, we obtain the following non-vanishing elements of its correlation tensor $T_{xx} = T_{yy} = -p$ and $T_{zz} = -1$.
Therefore, the decision tree allows detection of entanglement in this class of states in two steps.
Note that the state is entangled already for an infinitesimal admixture of the Bell singlet state.
We also verified numerically that for a hundred random choices of local coordinate systems, the decision tree detects entanglement even for $p>10^{-3}$.

\begin{figure}
\includegraphics[width=0.40\textwidth]{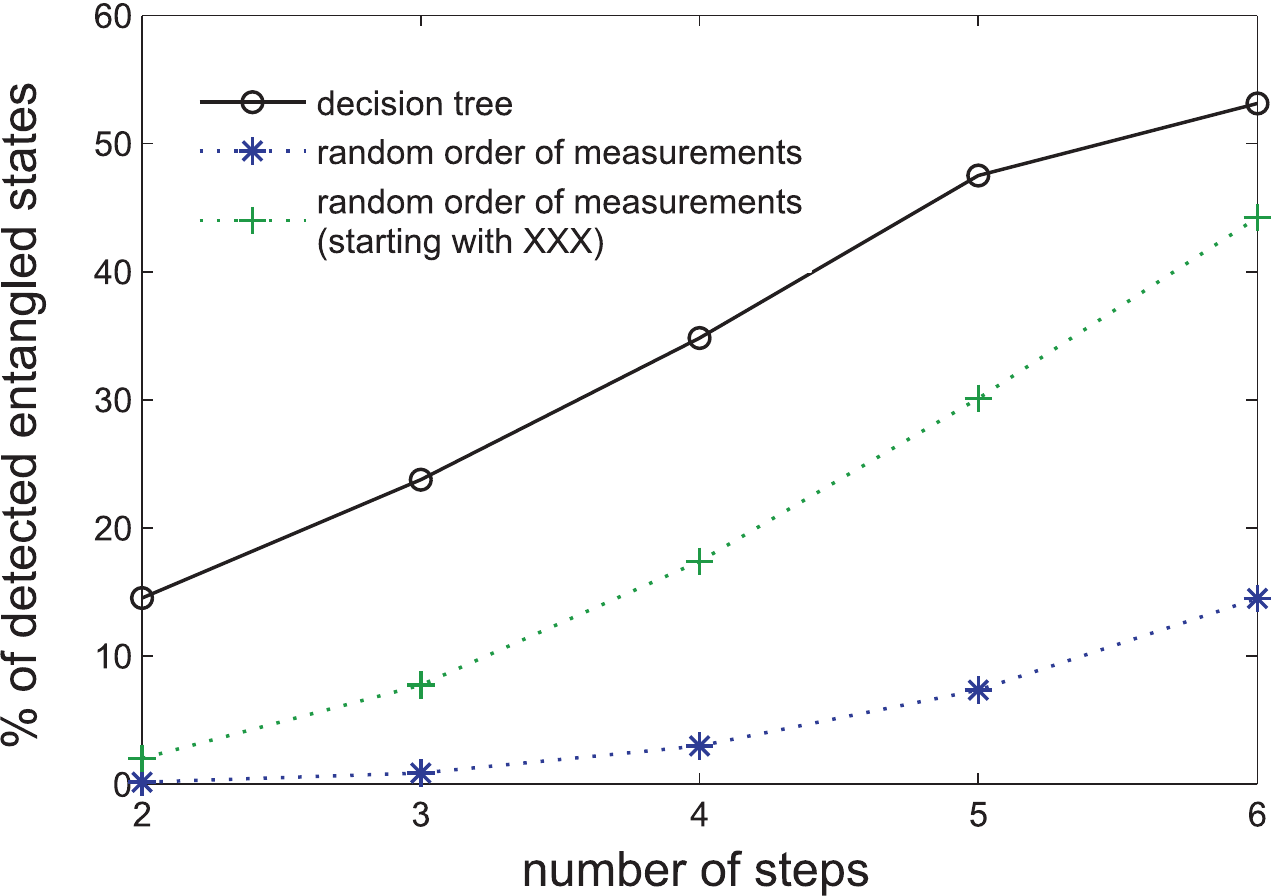}
\caption{Efficiency of one branch of the decision tree for three qubit random pure states.
The states were uniformly sampled according to the Haar measure.}
\label{fig:DT_3}
\end{figure}

\subsubsection{Three and more qubits}
\label{SEC_3Q}

\begin{figure}
\includegraphics[width=0.40\textwidth]{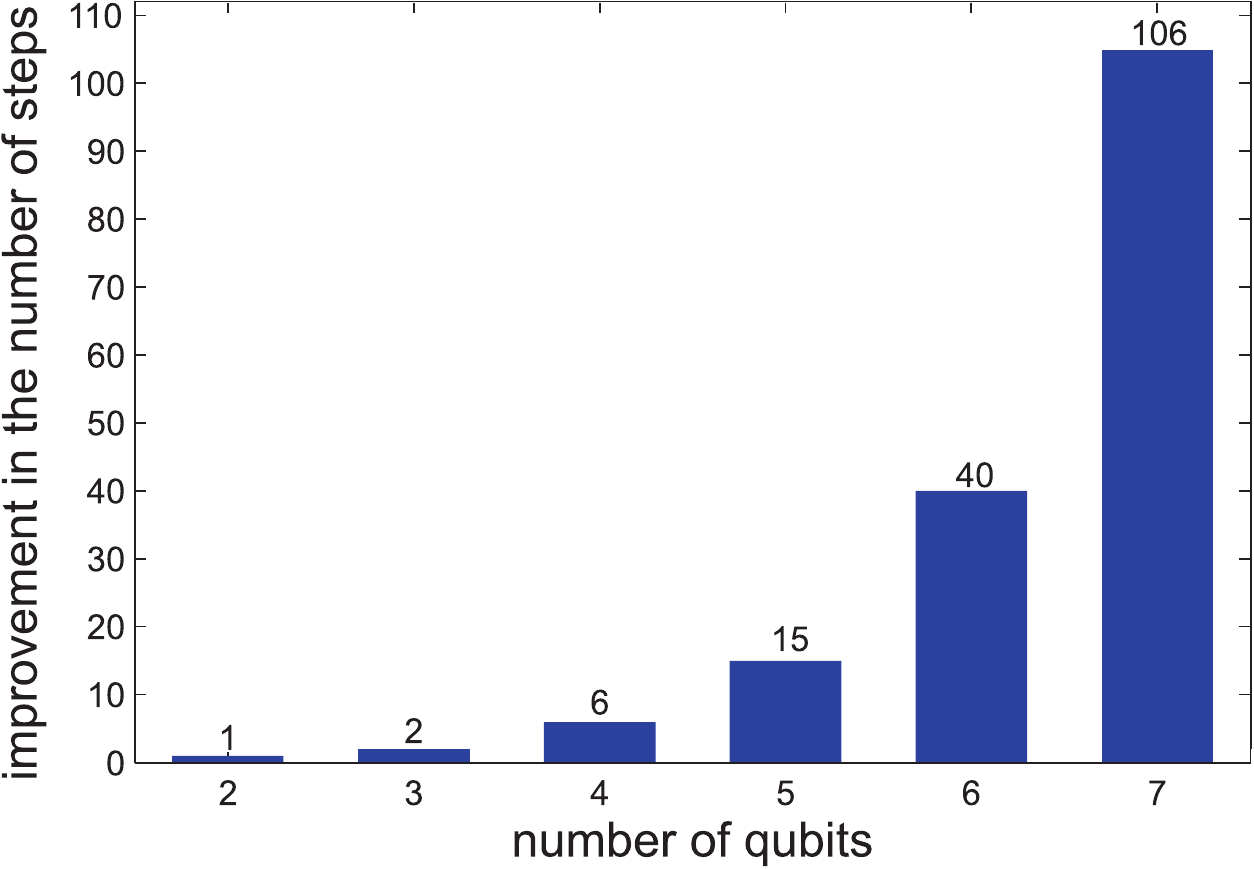}
\caption{Efficiency of one branch of the decision tree for many qubits compared with random choice of measurements. The plot shows the gain in the number of measurements provided by the decision tree. Pure states were uniformly sampled according to the Haar measure and the percentage of detected entangled states was calculated for different number of steps (measurements) in the tree as well as for the random order of measurements that start with $X \otimes \dots \otimes X$ for a fair comparison. We then compare the number of measurements for which the percentage of detected entangled states using the decision tree is the same as using the randomized measurements and plot here the maximal difference between them. The improvement provided by the tree grows exponentially with the number of qubits.}
\label{fig:DT_7}
\end{figure}

Similarly to the two-qubit case we also studied the efficiency of the three-qubit decision tree of Fig. \ref{DT_3QUBITS} as well as similar trees for higher number of qubits.
The results for three qubits are presented in Fig. \ref{fig:DT_3} and reveal that the decision tree is roughly two steps ahead of the protocol with random order of measurements for small number of steps. In general, the number of steps the decision tree is ahead of the protocol with random order of measurements grows exponentially with the number of qubits (see Fig. \ref{fig:DT_7}). The intuition behind is that once big correlations are measured using the decision tree, a set of measurements exponential in size is excluded whereas these measurements would still be randomly sampled in the other protocol.

\section{Experiments}

\begin{figure}
\includegraphics[width=0.49\textwidth]{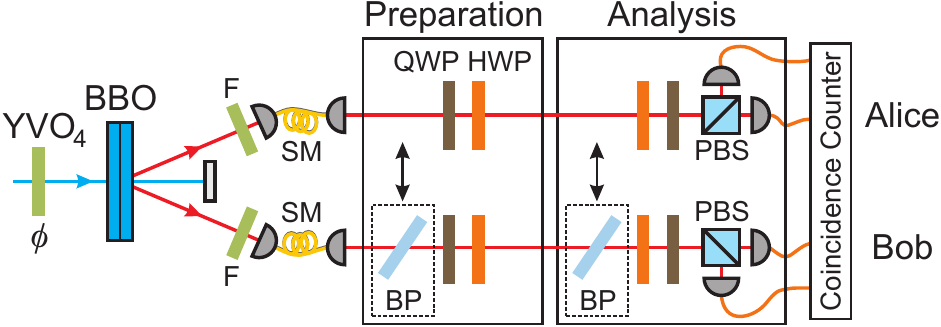}
\caption{Scheme of the experimental type~I SPDC source used to prepare the state $\frac{1}{\sqrt{2}}(\ket{HH}+\mathrm{e}^{i\phi}\ket{VV})$. The phase $\phi$ can be set by an yttrium vanadate crystal (YVO$_{4}$). Spectral filtering is performed by means of interference filters (F) and spatial filtering by single mode fibers (SM). Half- (HWP) and quarter-waveplates (QWP) are used for state preparation and analysis. Brewster plates (BP) enable performing the filter operation and the preparation of asymmetric states.}
\label{source2}
\end{figure}

The entanglement detection schemes introduced above are experimentally evaluated by analyzing a variety of multi-qubit entangled states. These states were created by spontaneous parametric down conversion (SPDC). Here, for the preparation of two qubit entangled states a type I source with two crossed optically contacted $\beta$-Barium-Borate (BBO) crystals of $1$mm thickness is used, see Fig.~\ref{source2}~\cite{KWIAT}. The computational basis $|0\rangle$ and $|1\rangle$ as introduced before is encoded in the polarization state $|H\rangle$ and $|V\rangle$, respectively. 
A continuous wave laser diode at $402$nm from NICHIA is used to pump the BBO crystals with approximately $60$mW. The polarization of the pump light is oriented at $45^{\circ}$  allowing to equally pump both crystals and to emit $HH$ and $VV$ polarized photon pairs with the same probability. However, a delay longer than the pump photon coherence length is acquired between the photon pairs generated in the first or second crystal over the length of the crystals, reducing their temporal indistinguishability. Therefore, an Yttrium-Vanadate (YVO$_{4}$) crystal of $200\mu m$ thickness is introduced in front of the BBOs to precompensate for the delay and to set the phase $\phi$ between $HH$ and $VV$. Using this configuration entangled states of the form $\ket{\Psi}=\frac{1}{\sqrt{2}}(\ket{H}\ket{H}+e^{i\phi}\ket{V}\ket{V})$ are generated~\cite{THESISPAVEL}.

In order to reduce the spectral bandwidth of the photon pairs, interference filters centered at $805$nm with a bandwidth of $7$nm are used. Spatial filtering is accomplished by coupling the photons at corresponding points of their emission cones into a pair of single mode fibers. Polarization controllers allow for the compensation of the polarization rotation of the fibers. Then, the photons are transmitted through a set of quarter (QWP)- and half (HWP)- waveplates allowing an arbitrary transformation of the polarization state in each path. A set of Brewster plates with a loss rate up to $\approx$60\% for $V$ and high transmission for $H$ polarized light can be introduced in front of the waveplates to enable preparation of states. For the analysis both Alice and Bob are provided with HWP and QWP as well as a filter (another Brewster plate) for Bob. Photons are then projected onto $\ket{H}$ and $\ket{V}$ implemented by a polarizing beamsplitter (PBS) and respective detectors. Note that local filtering can 
also be accomplished by a polarizer. The output modes of the analyzing PBS are coupled into multimode fibers connected to avalanche photon detectors (SPCM-AQ4C Perkin-Elmer module) with a photon detection efficiency of $\approx 50\%$. 
A coincidence logic is applied to extract the respective coincidence count rates within a time accuracy $<10ns$. 
The observed coincidence rate is approximately 200s$^{-1}$ and a measurement time of 10s per basis setting allows to register about 2000 events.
\newline

\subsection{Schmidt Decomposition}

\begin{figure}
\includegraphics[width=0.4\textwidth]{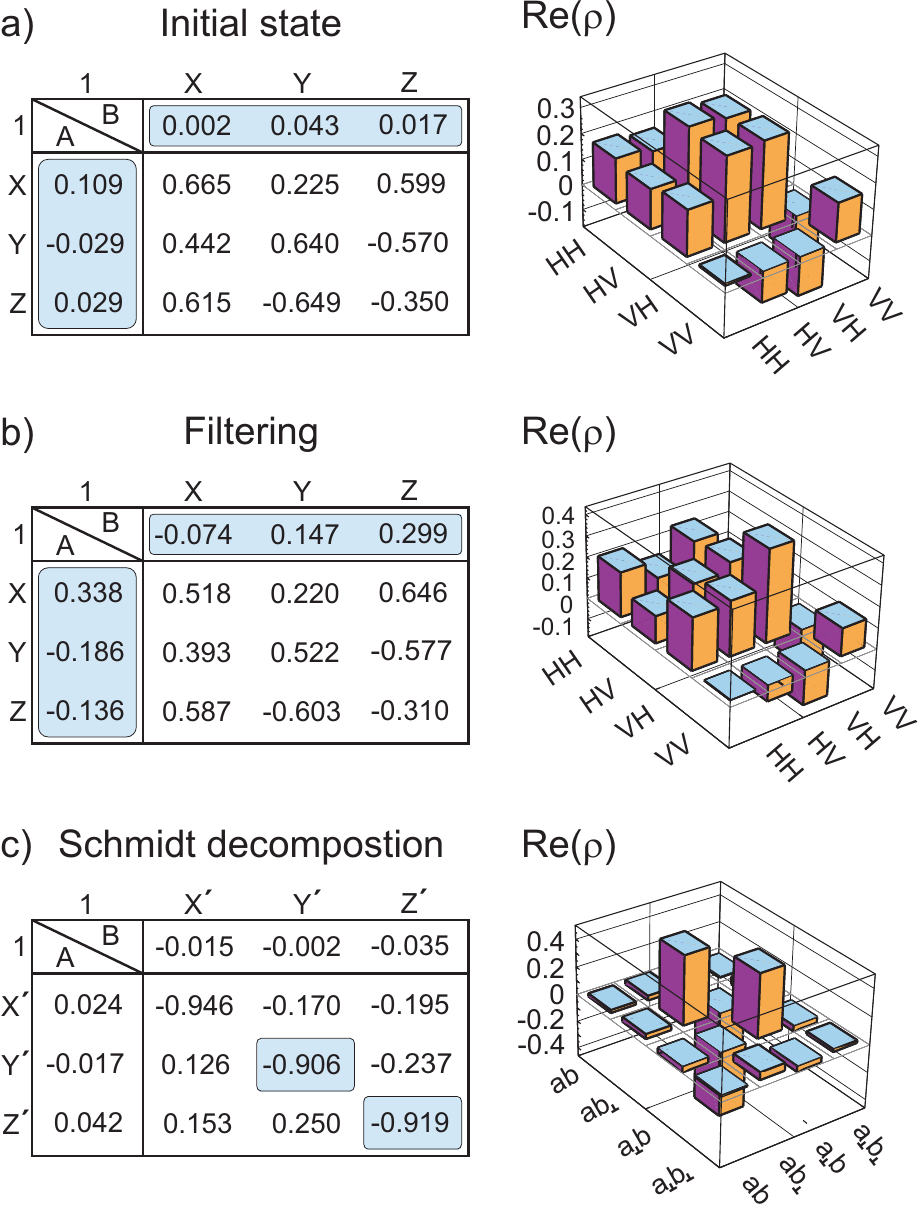}
\caption{Schmidt decomposition of a maximally entangled unknown state. The correlation tensor and the density matrix are determined a) before and b) after applying local filtering. After removing the filter, the state can be measured in its Schmidt basis c).}
\label{FIG_S}
\end{figure}

\begin{figure}
\includegraphics[width=0.3\textwidth]{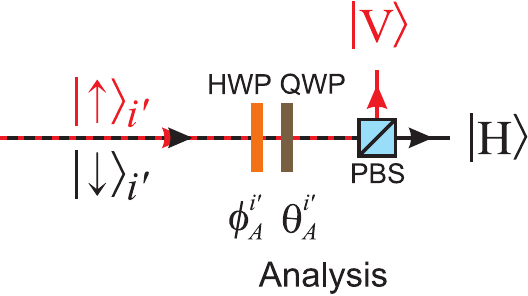}
\caption{If Alice wants to measure in the basis $\sigma_{i'} = \ket{\downarrow}_{i{'}}\bra{\downarrow}_{i{'}} - \ket{\uparrow}_{i{'}}\bra{\uparrow}_{i{'}}$ ($i=x,y,x$) the HWP and the QWP of the polarization analysis have to be aligned such that $\ket{\downarrow}_{i{'}}$ and $\ket{\uparrow}_{i{'}}$ are detected at different outputs of the PBS. The same holds for Bob.}
\label{FIG_SchmidtPol}
\end{figure}

In order to perform the measurement in the Schmidt basis we first have to determine basis vectors from the Bloch vectors observed by Alice and Bob. Let us consider
the state depicted in Fig.~\ref{FIG_S}a. The table to the left shows the correlation tensor elements $T_{ij}$ with the Bloch vectors of Alice (Bob) in the left most column (top row). 
For the application of the Schmidt decomposition method, Alice and Bob measure first their respective Bloch vectors (measurements actually to be performed are indicated by the blue shaded fields). Since they are close to 0,
$\vec{T}_{A}=(0.002,0.043,0.017)$ and $\vec{T}_{B}=(0.109,-0.029,0.029)$, the
next step of the algorithm is to apply local filtering as described by the scheme of Fig.~\ref{FIG_ALGO}. The
filtered state shown in Fig.~\ref{FIG_S}b has non-vanishing Bloch vectors,
$\vec{T}_{A}=(0.338,-0.186,-0.136)$ and ${\vec{T}}_{B}=(-0.074,0.147,0.299)$,
which can be used to find the corresponding Schmidt basis of the shared
two-qubit state\footnote{It is to note that the Bloch vectors are determined from coincidence measurements. This is due to the low detection efficiency of correlated photons.}. If the phase $\phi$ is not determined  by an additional
correlation measurement there are infinitely many such bases. As shown in section \ref{localresults} one possible choice is to redefine the local basis of Alice and Bob according to equations (\ref{new_directions}).
Measuring along $\sigma_{i{'}}$ corresponds to a projection on its eigenstates $\ket{\downarrow}_{i{'}}$ and $\ket{\uparrow}_{i{'}}$.
The task now is to find the angles for the waveplates of Alice and Bob $\theta_A^{i'}/\phi_A^{i'}$ and $\theta_B^{i'}/\phi_B^{i'}$, respectively. Since the PBS of the polarization analysis shown in Fig.~\ref{FIG_SchmidtPol} always projects on $\ket{H}$ and $\ket{V}$, the angles are calculated under the condition that $\ket{\downarrow}_{i{'}}$ ($\ket{\uparrow}_{i{'}}$) is rotated, up to a global phase $\tau$, to $\ket{H}$ ($\ket{V}$), e.g. for Alice
\begin{eqnarray}
U_\mathrm{QWP}(\theta_A^{i'})U_\mathrm{HWP}(\phi_A^{i'})\ket{\downarrow}_{i{'}} &=& \mathrm{e}^{i \tau_1}\ket{H},
\label{eq21}\\
U_\mathrm{QWP}(\theta_A^{i'})U_\mathrm{HWP}(\phi_A^{i'})\ket{\uparrow}_{i{'}} &=& \mathrm{e}^{i \tau_2}\ket{V},
\end{eqnarray}
where $U$ labels the unitary operation of the corresponding waveplate. The angles $\theta_A^{i'}$/$\phi_A^{i'}$ and $\theta_B^{i'}$/$\phi_B^{i'}$ can be found by (numerically) solving the equation
\begin{eqnarray}
|\bra{H}U_\mathrm{QWP}(\theta_A^{i'})U_\mathrm{HWP}(\phi_A^{i'})\ket{\downarrow}_{i{'}}|^2 &=& 1,
\label{eq26}
\end{eqnarray}
and similarly for Bob.
Using this scheme, we find the angles for Alice's and Bob's waveplates, such that their qubits are measured in the primed bases, presented in Table \ref{settingsME}.
\begin{center}
\begin{table}[b]
\begin{tabular}[b]{c c c} 
\multicolumn{3}{c}{Alice} \\
\hline \hline 
 & $\frac{\lambda}{2}$ & $\frac{\lambda}{4}$\\  \hline 
$\sigma_{x'}$ & $22.6^\circ$ & $25.8^\circ$ \\
$\sigma_{y'}$ & $-15.8^\circ$ & $13.3^\circ$ \\
$\sigma_{z'}$ & $-9.9^\circ$ & $-12.8^\circ$ \\ \hline \hline
\end{tabular}
\quad
\begin{tabular}[b]{c c c} 
\multicolumn{3}{c}{Bob} \\
\hline \hline 
 & $\frac{\lambda}{2}$ & $\frac{\lambda}{4}$\\  \hline 
$\sigma_{x'}$ & $6.6^\circ$ & $4.6^\circ$ \\
$\sigma_{y'}$ & $7.2^\circ$ & $-30.6^\circ$ \\
$\sigma_{z'}$ & $34.7^\circ$ & $13.5^\circ$ \\\hline \hline
\end{tabular}
\caption{Waveplate settings for Alice and Bob to measure the maximally entangled state shown in Fig.~\ref{FIG_S}a in the Schmidt basis and the complementary directions.}
\label{settingbsME}
\end{table}
\end{center}

After removing the filter, Alice and Bob can now measure in their new bases and reveal entanglement by measuring $T_{z{'}z{'}}$ followed by $T_{y{'}y{'}}$ and possibly $T_{y{'}x{'}}$. Here the two measurements suffice to reveal entanglement as $T_{z{'}z{'}}^2+T_{y{'}y{'}}^2=1.665\pm 0.05>1$.\\

\begin{figure}
\includegraphics[width=0.4\textwidth]{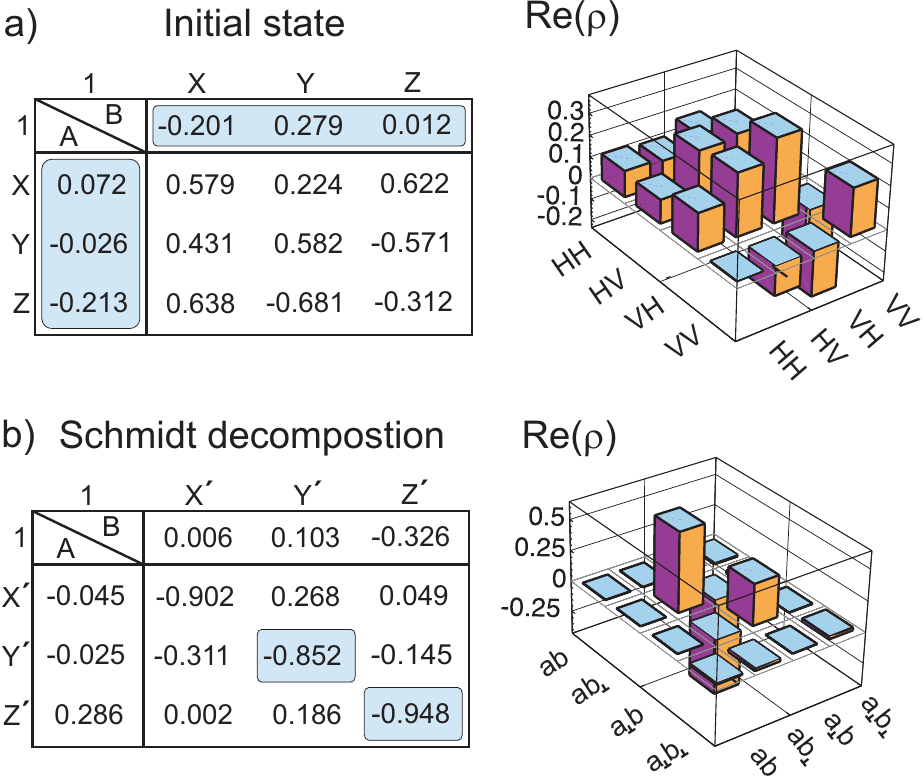}
\caption{Schmidt decomposition of a non-maximally entangled state. The correlation tensor and density matrix is displayed for an a) unknown asymmetric state. 
The state has nonzero Bloch vectors enabling to determine the corresponding Schmidt basis for which the measured correlations are maximal b).}
\label{FIG_SA}
\end{figure}

In full analogy to the previous example, it is also possible to apply the Schmidt decomposition scheme to a non-maximally entangled state, e.g., as presented in Fig.~\ref{FIG_SA}a. For using the Schmidt decomposition strategy, first both parties agree on measuring their respective Bloch vectors $\vec{T}_{A}=(0.072,-0.026,-0.213)$ and $\vec{T}_{B}=(-0.201,0.279,0.012)$.
As they already can be distinguished from noise, Alice and Bob can find the Schmidt bases without applying the filter operation. The angle settings of the waveplates for analyzing in the Schmidt bases are again calculated using (\ref{eq26}) and are shown in Table \ref{settingsAsym}. Again, the state is proved to be entangled after only two correlation measurements since $T_{z{'}z{'}}^2+T_{y{'}y{'}}^2=1.624\pm 0.047>1$, see Fig~\ref{FIG_SA}b.

\begin{center}
\begin{table}[b]
\begin{tabular}[b]{c c c} 
\multicolumn{3}{c}{Alice} \\
\hline \hline 
 & $\frac{\lambda}{2}$ & $\frac{\lambda}{4}$\\  \hline 
$\sigma_{x'}$ & $0.7^\circ$ & $0.8^\circ$ \\
$\sigma_{y'}$ & $-13.5^\circ$ & $17.9^\circ$ \\
$\sigma_{z'}$ & $-35.2^\circ$ & $-27.0^\circ$ \\\hline \hline
\end{tabular}
\quad
\begin{tabular}[b]{c c c} 
\multicolumn{3}{c}{Bob} \\
\hline \hline 
 & $\frac{\lambda}{2}$ & $\frac{\lambda}{4}$\\  \hline 
$\sigma_{x'}$ & $21.9^\circ$ & $9.4^\circ$ \\
$\sigma_{y'}$ & $40.0^\circ$ & $-55.0^\circ$ \\
$\sigma_{z'}$ & $42.0^\circ$ & $3.3^\circ$ \\\hline \hline
\end{tabular}

\caption{Waveplate settings for Alice and Bob to measure the asymmetric state shown in Fig.~\ref{FIG_SA}a in the Schmidt basis and the complementary directions.}
\label{settingsAsym}
\end{table}
\end{center}

\subsection{Decision tree}

\subsubsection{Two qubits}

\begin{figure*}
\includegraphics[width=0.95\textwidth]{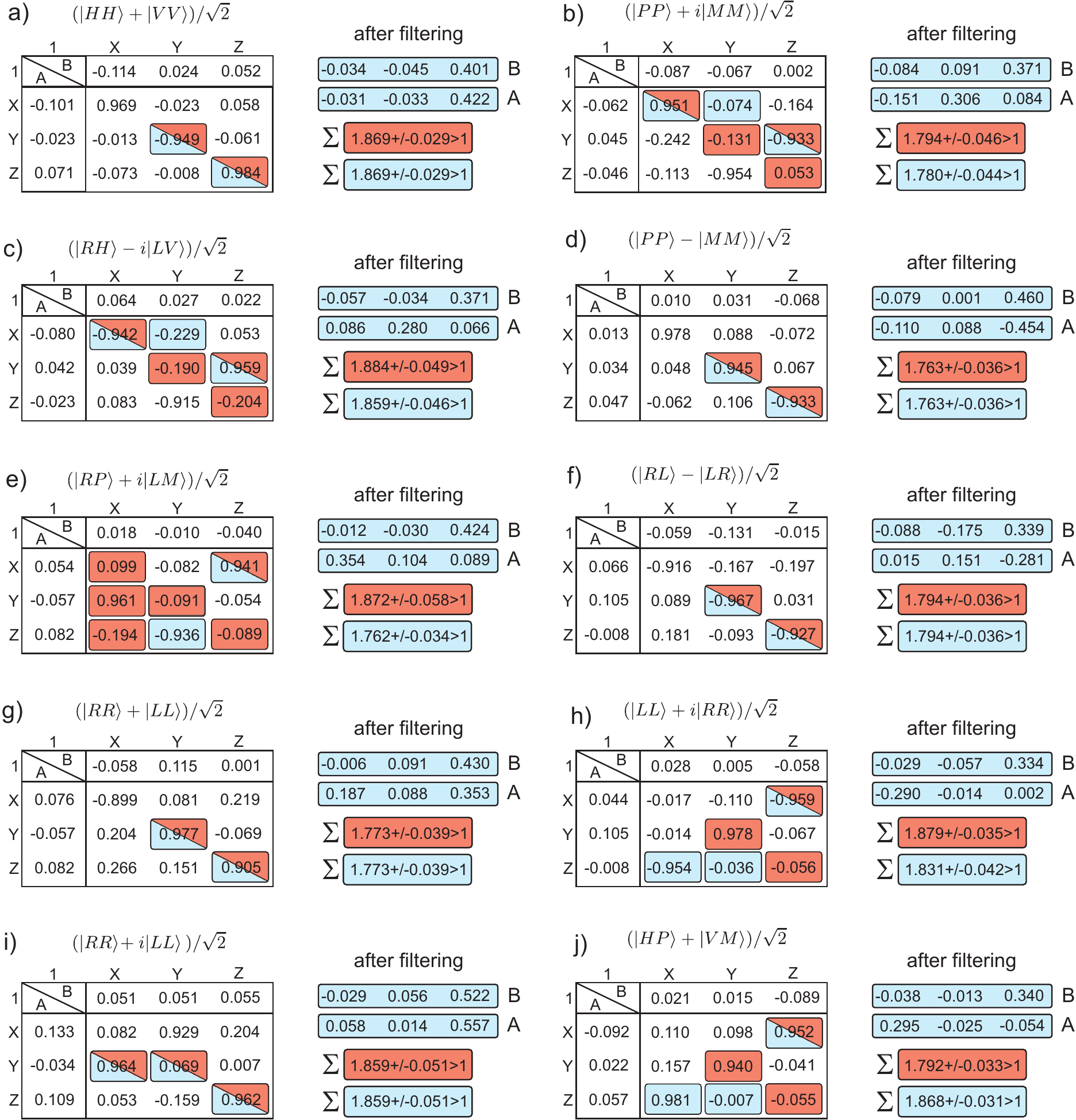}
\caption{Application of the decision tree on a selection of maximally entangled states, allowing to determine the entanglement of the state by measuring the correlations marked in red color. As an alternative, local filtering is applied in order to extract the correlations with maximal value (blue correlations).
}
\label{FIG_TDTSym}
\end{figure*}

\begin{figure*}
\includegraphics[width=0.95\textwidth]{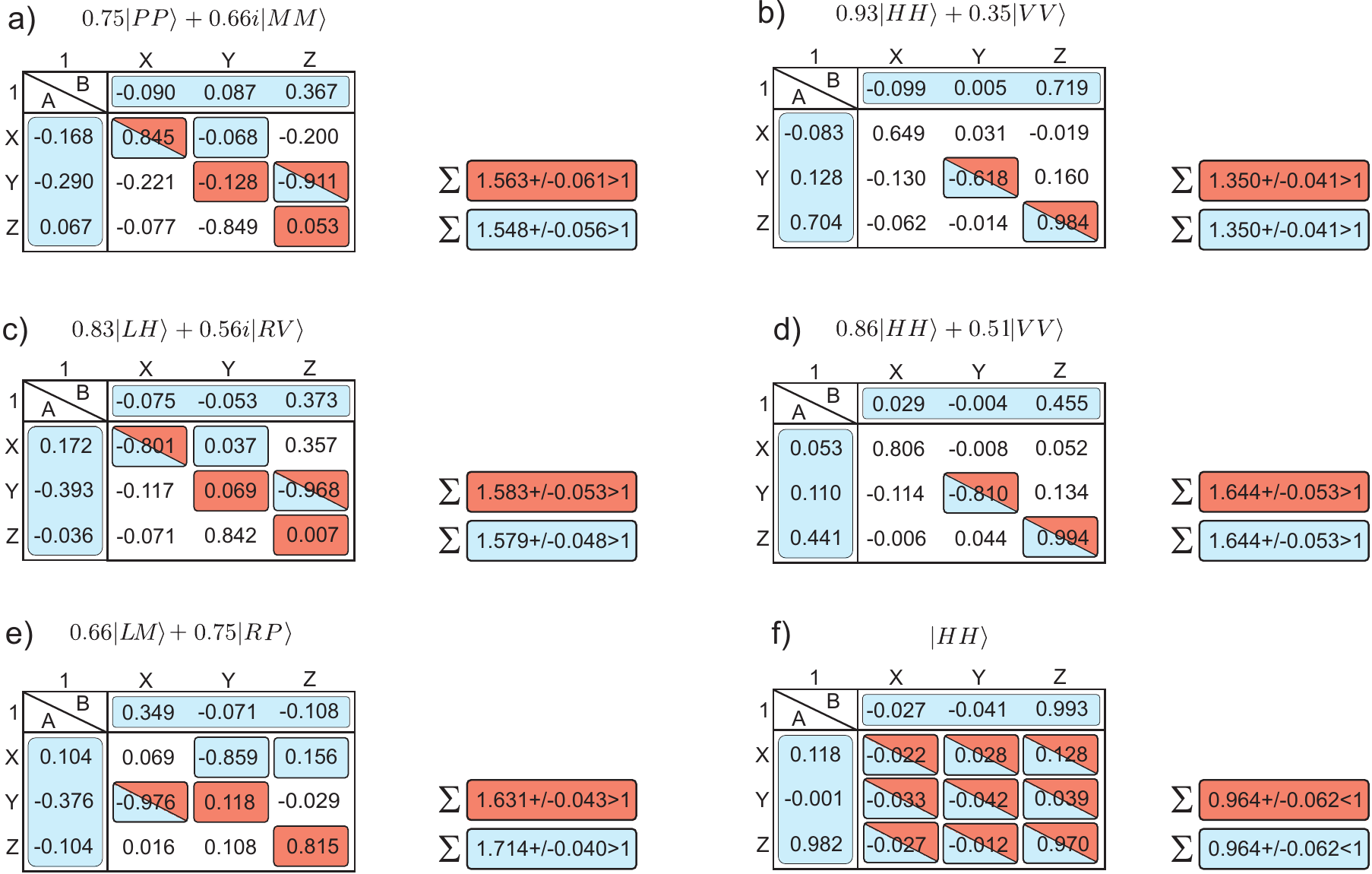}
\caption{Application of the decision tree on a selection of non-maximally entangled states, allowing to determine the entanglement of the state by measuring the correlations marked in red color. Due to the asymmetry of the states, local filtering is unnecessary, and the information on the Bloch vectors can be used to detect entanglement with a maximal number of $3$ correlation measurements (blue correlations). Panel f) shows that for a product state full set of correlations does not reveal entanglement, as it should be.}
\label{FIG_TDTAsym}
\end{figure*}

Let us first consider the two states analyzed above using Schmidt decomposition. For the first state (Fig.~\ref{FIG_S}) we see that a direct application of the decision tree shown in Fig.~\ref{FIG_DT} would require four correlation measurements to reveal entanglement, namely $T_{zz}^2+T_{yy}^2+T_{xz}^2+T_{zx}^2=(-0.350)^2+0.640^2+0.599^2+0.615^2=1.33\pm 0.03> 1$. Similarly, the analysis of the second state (Fig.~\ref{FIG_SA}) would require four correlation measurements to determine entanglement, namely $T_{zz}^2+T_{yy}^2+T_{xx}^2+T_{xz}^2=(-0.312)^2+0.582^2+0.579^2+0.622^2=1.158 \pm 0.030>1$. This shows that quite a few more correlation measurements are needed when using the decision tree. Yet, it saves measuring the Bloch vectors and filtering operations. 
To illustrate the entanglement detection scheme we further apply it to a selection of maximally entangled states (Fig.~\ref{FIG_TDTSym}) and to non-maximally entangled states (Fig.~\ref{FIG_TDTAsym}).

For didactical reasons, the full correlation tensors are depicted, in both cases. In order to reveal entanglement, the decision tree requires the measurement of a number of correlations much smaller than needed to reconstruct the full density matrix. Following the lines as described in section \ref{section_DT}, only correlation measurements shaded red are required to detect entanglement. As an example let us consider the state $\frac{1}{\sqrt{2}}(|RR\rangle + |LL\rangle)$ (Fig.~\ref{FIG_TDTSym}g), for which a measurement of the two correlations $T_{zz}=0.905$ and $T_{yy}=0.977$ suffices to reveal entanglement since $T_{zz}^2+T_{yy}^2=1.773\pm0.039>1$. In contrast, for the state $\frac{1}{\sqrt{2}}(|RP\rangle + i|LM\rangle)$ (Fig.~\ref{FIG_TDTSym}e) the algorithm only stops after six steps as the measurements of $T_{zz}=-0.089$, $T_{yy}=-0.091$, $T_{xx}=0.099$, $T_{zx}=-0.194$, $T_{xz}=0.941$ and $T_{yx}=0.961$ are required to beat the threshold, i.e. $1.872\pm0.058>1$. A similar reasoning is applied to reveal 
entanglement of other two-qubit states.

The entanglement detection scheme is further applied to a selection of non-maximally entangled states (Fig.~\ref{FIG_TDTAsym}). As an example, let us consider the state $0.83|LH\rangle + 0.56 i|RV\rangle$ (Fig.~\ref{FIG_TDTAsym}c), for which our method reveals entanglement after four steps, as the measurements of $T_{zz}=0.007$, $T_{yy}=0.069$, $T_{xx}=-0.801$ and $T_{yz}=-0.968$ give a value of $1.583\pm0.067>1$. Similarly, as expected, for a separable state such as $|HH\rangle$ (Fig.~\ref{FIG_TDTAsym}f), our entanglement criterion delivers a value of $\sum_{k,l=1}^{3} T_{kl}^{2}=0.964\pm0.062<1$ for measuring all correlations, not revealing entanglement clearly. These states of course can be analyzed also using Schmidt decomposition.
For maximally entangled states, the Bloch vectors after local filtering are also shown (blue color, see Fig.~\ref{FIG_TDTSym}), while for non-maximally entangled states (Fig.~\ref{FIG_TDTAsym}) no local filtering is required since the Bloch vectors are already non-vanishing. In all cases, only one entry of the respective Bloch vectors is large compared to the others. Therefore, no realignment of the analyzers is necessary. Due to Schmidt decomposition the decision tree should start with a correlation measurement along a direction in which we see a big local expectation value. In such a case it is sufficient to cyclically relabel the required measurements as defined for the original decision tree. Following this method, it is possible to detect entanglement with a maximum number of three steps. The first correlation to be measured is determined by the Bloch vectors after applying local filtering. 
%

\subsubsection{Many qubits}

\label{ManyQubits}

For the demonstration of multi-qubit entanglement detection, we use a family of three-photon polarization entangled Gda\'nsk (G) states~\cite{GDANSK} and the four-qubit Dicke state.
The G states are defined by
\begin{equation}
| G (\alpha) \rangle = \cos(\alpha) \ket{W} + \sin(\alpha) | \overline{W} \rangle,
\label{G_STATE}
\end{equation}
where $\ket{W} = \frac{1}{\sqrt{3}}(\ket{HHV}+\ket{HVH}+\ket{VHH})$, and in order to obtain $| \overline{W} \rangle$ one exchanges $H$ and $V$.
The four-qubit Dicke state with two ``excitations'' reads
\begin{eqnarray}
| D_4^{(2)} \rangle & = & \frac{1}{\sqrt{6}} (\ket{HHVV}+\ket{HVHV}+\ket{VHHV} \nonumber \\
& + & \ket{HVVH}+\ket{VHVH}+\ket{VVHH}).
\label{DICKE_STATE}
\end{eqnarray}

{\em Generalized three-qubit G state.} In order to observe these states, a collinear type II SPDC source together with a linear setup to prepare the four-photon Dicke state $D_4^{(2)}$ is used \cite{KRISCHEK, KIESEL}.
The three-photon state is obtained if the first photon is measured to be $\cos(\alpha) \ket{H} + \sin(\alpha) \ket{V}$ polarized.\\

The protocol for entanglement detection starts with observers locally measuring the polarization of their respective photons enabling them to individually determine the Bloch vectors. 
\begin{itemize}
\item For the $G(\pi / 4)$ state we obtain 
$T_{i00}=(0.636,-0.008, -0.015)$, 
$T_{0j0}=(0.623,-0.092,0.010)$ and
$T_{00k}=(0.636, 0.070,0.022)$. 
The Bloch vectors suggest that the correlation $T_{xxx}$ is big. Therefore the decision tree starts with the measurement of $T_{xxx} =0.904 \pm 0.025$ 
and continues with $T_{xzz} = -0.578 \pm 0.025$ (see Fig.~\ref{DT_3QUBITS}). These two measurements already prove entanglement because $T_{xxx}^2 + T_{xzz}^2 = 1.152 \pm 0.038 > 1$.

\item For the $W$ state, $G(\pi / 2)$, the Bloch vectors are 
$T_{i00}=(0.016, -0.070, 0.318)$, 
$T_{0j0}=(-0.010, -0.073, 0.308)$ and
$T_{00k}=(-0.011, -0.0547, 0.319)$, which suggest that now the correlation $T_{zzz}$ is big. Indeed, we observe $T_{zzz} = -0.882 \pm 0.025$.
The decision tree is the same as above but with local axes renamed as follows $x \to z \to y \to x$.
Therefore, the second measurement has to be $T_{zyy}$. With $T_{zyy} = 0.571 \pm 0.025 $ we again prove entanglement as $T_{xxx}^2 + T_{zyy}^2 = 1.104 \pm 0.037 > 1$.

%
\end{itemize}

{\em Four-qubit Dicke state.}
Here, we have vanishing Bloch vectors, 
$T_{i000}=(-0.020, -0.016, 0.007)$,
$T_{0j00}=(-0.011, -0.029, 0.014)$,
$T_{00k0}=(-0.018, -0.020, -0.004)$ and
$T_{000l}=(-0.009, -0.022, 0.008)$.
We construct a set of mutually commuting operators which form the first branch of the four qubit decision tree starting with $T_{zzzz}$,
$\{ zzzz \rightarrow zzxx \rightarrow zxzx \rightarrow zxxz \rightarrow xzxz \rightarrow xxzz \rightarrow xzzx \rightarrow xxxx \rightarrow yyyy \}$. After measuring the correlations
$T_{zzzz}=0.848 \pm 0.025$
$T_{zzxx}=-0.533\pm 0.025$
$T_{zxzx}=-0.552\pm 0.025$ our algorithm succeeds since $T_{zzzz}^2 + T_{zzxx}^2 + T_{zxzx}^2  = 1.3082 \pm 0.041 > 1$.

\section{Conclusions}

The entanglement of arbitrary multi-qubit states can be efficiently detected based on two methods described here. Both methods employ a criterion based on the sum of squared correlations. Combining this with an adaptive determination of the correlations to be measured allows to succeed much faster than standard tomographic schemes. The first one, particularly designed for two-qubit states determines the Schmidt decomposition from local measurements only, where at most three correlation measurements are sufficient for entanglement detection. The second one employs a decision tree to speed up the analysis.
Its design is based on correlation complementarity and prevents one from measuring less informative correlations. The performance of the scheme is numerically analyzed for arbitrary pure states, and in the two-qubit case, for mixed states. The schemes succeed on average at least one step earlier as compared with random sampling on two qubit states, with an exponentially increasing speedup for a higher number of qubits.
Our results encourage the application of these schemes in state of the art experiments with quantum states of increasing complexity.


\emph{Acknowledgments.}---
We thank the EU-BMBF project QUASAR and the EU projects QWAD and QOLAPS for supporting this work.
TP acknowledges support by the National Research Foundation, the Ministry of Education of Singapore, start-up grant of the Nanyang Technological University, and NCN Grant No. 2012/05/E/ST2/02352.
CS thanks QCCC of the Elite Network of Bavaria for support.

\end{document}